\documentclass[conference,10pt]{IEEEtran}
\IEEEoverridecommandlockouts
\usepackage{amsmath,amssymb,bm,mathtools}
\usepackage{graphicx}
\usepackage{booktabs,multirow,array}
\usepackage{cite}
\usepackage{amsthm}
\usepackage{caption}
\DeclareCaptionLabelSeparator{periodspace}{.~}
\usepackage{float}
\usepackage{subcaption}
\usepackage{pgfplots}
\pgfplotsset{compat=1.18}
\usepgfplotslibrary{groupplots}
\usepackage[hidelinks]{hyperref}
\newtheorem{prop}{Proposition}

\newcommand{\C}{\mathbb C}
\newcommand{\ip}[2]{\langle #1,#2\rangle}
\newcommand{\mx}[1]{\mathbf{#1}}
\newcommand{\tr}{\mathrm{tr}}
\newcommand{\vecop}{\mathrm{vec}}
\newcommand{\uvar}{u}
\newcommand{\thvar}{\theta}
\newcommand{\gf}[1]{\textcolor{black}{{#1}}}
\newcommand{\bs}[1]{\boldsymbol{#1}}

\title{Living Off the Grid: Continuous Range-Angle Super-Resolution for Near-Field XL-MIMO
\thanks{This work is supported in part by Digital Futures and the EU Horizon projects MULTIPLY-6G and QUEST-6G under Project IDs 101293106 and 101292676, respectively. G. Fodor was also supported by SSF grant FUS21-0004 SAICOM.}

}

\author{
Sajad Daei$^{1}$, G\'abor Fodor$^{1,2}$, Mikael Skoglund$^{1}$\\
$^{1}$KTH Royal Institute of Technology, Stockholm, Sweden\\
$^{2}$Ericsson Research, Stockholm, Sweden\\
\{sajado,gaborf,skoglund\}@kth.se
}

\begin{document}
\maketitle

\begin{abstract}
Near-field extremely large multiple-input multiple-output (XL-MIMO) breaks the assumptions that make classical super-resolution effective: the receiver acquires only a limited set of compressed pilot observations, while each propagation path is jointly determined by angle and distance under a spherical-wave model. This invalidates the far-field Vandermonde structure exploited by conventional methods, while many existing polar-domain formulations rely on range-angle discretization or partial off-grid refinement, thereby inheriting mismatch, coherence, and resolution tradeoffs.
This paper develops a continuous 2D super-resolution framework for hybrid near-field measurements that avoids range and angle gridding. The key idea is to reparameterize distance through inverse range, which reveals a compact spectral structure for the near-field spherical-wave manifold. Building on this observation, we introduce a panelized weighted fitting strategy that converts the range-dependent Fresnel terms into a stable transform-domain representation, resulting in a lifted model in which each continuous range-angle pair is embedded as a structured rank-one atom and the measurement model remains linear under hybrid combining. Recovery is then posed as a 2D atomic norm minimization, with path localization certified through a dual polynomial over the transformed domain.
Numerical experiments demonstrate accurate support recovery in a representative noiseless setting using only a few compressed hybrid measurements. These results establish the proposed inverse-range atomic norm viewpoint as a new gridless foundation for near-field sensing and channel estimation in hybrid XL-MIMO and integrated sensing and communication systems.
\end{abstract}

\section{Introduction}
For decades, multi-antenna signal processing has benefited from a simple geometric approximation: across the array, wavefronts are nearly planar. This approximation underlies the Fourier/Vandermonde structure behind beamforming, direction finding, and classical super-resolution, and it is what allows angle estimation to be treated as a one-dimensional spectral problem\cite{tang2013compressed,candes2014towards,daei2025when,daei2025near}.
That approximation
% \gf{can be substantially improved in extremely large (XL-MIMO) systems.}
is no longer enough.
Extremely large multiple-input multiple-output (XL-MIMO) aperture arrays, millimeter-wave and sub-THz carriers, and the rise of integrated sensing and communication (ISAC) are pushing wireless systems into the near field. In this regime, propagation is governed by spherical rather than planar wavefronts, and each path is described jointly by angle and distance. The phase profile across the array therefore depends on both variables, so the far-field spectral model no longer applies directly \cite{liu2024near,daei2025near,daei_globecom}.
The difficulty is amplified by hardware. In hybrid XL-MIMO architectures, a base station may employ hundreds of antennas but only a small number of RF chains, so the channel is observed only through a limited set of compressed pilot measurements \cite{alkhateeb2014channel, cui2022channel,zhang2023near}. Near-field estimation is thus shaped by two simultaneous constraints: the underlying geometry is richer, while the available observations are fewer.

The standard response is to discretize angle and range and solve a sparse recovery problem over a polar-domain dictionary \cite{cui2022channel,zhang2023near,wang2023near,liu2024near}. Such methods are practical and can be effective for channel reconstruction. However, in hybrid XL-MIMO with few compressed measurements, grid refinement increases dictionary size and coherence, while coarse grids introduce basis mismatch\cite{chi2011sensitivity,tang2013compressed}. Moreover, in ISAC and geometry-aware communication, the continuous range-angle support is itself useful. This motivates a geometry-aligned formulation in which range and angle remain continuous physical parameters throughout the recovery process, rather than artifacts of a discretized polar dictionary.

Unlike polar-domain CS and off-grid refinement methods, the proposed approach does not start from a fixed range-angle dictionary. Unlike semi-gridless near-field formulations \cite{daei2026convexity,xi2025near}, it treats both range and angle continuously through the inverse-range lifted atom. Thus, the main novelty is not only the use of atomic norm minimization, but the inverse-range Bessel-Vandermonde lifting that converts the near-field spherical-wave model into a linear continuous 2D recovery problem under hybrid measurements.
This paper takes a different view. Rather than asking how to discretize range more finely, we ask a more basic question: \gf{\emph{are we even looking at range in the right coordinate system?}} The key observation is that near-field curvature, although it breaks the far-field model, does not destroy harmonic structure altogether. Instead, it enriches it. The angular dependence still carries a hidden Vandermonde core, while the effect of range is absorbed into curvature-induced coefficient functions. Once distance is viewed through inverse range rather than range itself, these coefficient functions become far more regular and amenable to compact spectral modeling.

This perspective leads to a Bessel-Vandermonde description of the near-field manifold. The spherical-wave steering vector can be interpreted as a harmonic expansion in which curvature introduces additional harmonics beyond the far-field model, while preserving a structured angular dependence. We exploit this structure by reparameterizing distance through inverse range and mapping it onto a periodic domain. In the transformed coordinate, the Fresnel/Bessel terms admit a compact spectral representation, yielding a model that is continuous in both variables: angle remains parameterized through its harmonic structure, while range is represented through a periodic inverse-range coordinate. Crucially, this transformed representation remains linear under hybrid combining, which makes it directly compatible with compressed pilot acquisition. Equivalently, each continuous range-angle pair is embedded as a structured rank-one atom in a higher-dimensional lifted space, and the near-field steering vector is represented as the action of a linear operator on this atom. This lifting step converts the original nonlinear parameter dependence into a linear inverse problem over a continuous atomic model.
\gf{The consequence of the above is a shift in viewpoint: Rather than approximating near-field propagation with a dense range-angle dictionary, we describe it through a small number of geometry-aligned spectral components.} 
This leads naturally to a 2D atomic norm formulation in which angle and range are both treated off the grid, and in which support recovery is certified through a dual polynomial over the transformed domain \cite{chandrasekaran2012convex,tang2013compressed,candes2014towards,sayyari2020blind,valiulahi2019two,safari2021off,maskan2023demixing}  . In hybrid near-field systems, where every measurement is valuable, such a representation offers both conceptual clarity and practical relevance.

The contributions of this work are listed below:
\begin{itemize}
\item \textbf{Hybrid XL-MIMO formulation for near-field super-resolution:}
We explicitly formulate the problem under a hybrid XL-MIMO architecture, where a base station (BS) with $N_r$ antennas but only $N_{\mathrm{RF}}\!\ll\! N_r$ RF chains acquires a small number of compressed pilot observations through analog combining. This places the proposed method in the practically relevant regime where near-field geometry is richer than in the far field, yet the available measurements are strongly limited.

\item \textbf{Continuous Bessel-Vandermonde lifting via inverse-range modeling:}
We develop a harmonic representation of the near-field steering vector that preserves a hidden Vandermonde structure in angle while absorbing the curvature-induced range dependence into a compact coefficient map. By reparameterizing distance through inverse range and approximating the resulting Fresnel/Bessel coefficient functions with a panelized weighted Fourier model, we obtain a lifted representation in which each continuous angle-range pair is embedded as a structured rank-one atom. This converts the original nonlinear near-field manifold into a linear operator acting on a continuous lifted variable.

\item \textbf{Gridless 2D recovery with convex certification:}
Building on the lifted model, we establish a fully gridless convex framework for joint angle-range super-resolution under hybrid measurements. We cast recovery as a 2D primal atomic norm minimization problem and derive a semidefinite representation of the primal problem through a multi-dimensional Toeplitz moment lift. The resulting continuous dual polynomial acts as an explicit certificate for support localization, enabling simultaneous off-grid recovery in angle and range without resorting to any discretization.
\item \textbf{Amplitude estimation via least-squares optimization:}
Once the continuous range-angle support has been identified, we estimate the corresponding complex path gains by solving a least-squares problem over the recovered atoms. This post-processing step provides the amplitude associated with each range-angle pair and completes the near-field channel reconstruction with low computational overhead.
\end{itemize}

\section{Hybrid XL-MIMO System Model}
\begin{figure}
    \centering
    \includegraphics[scale=0.3]{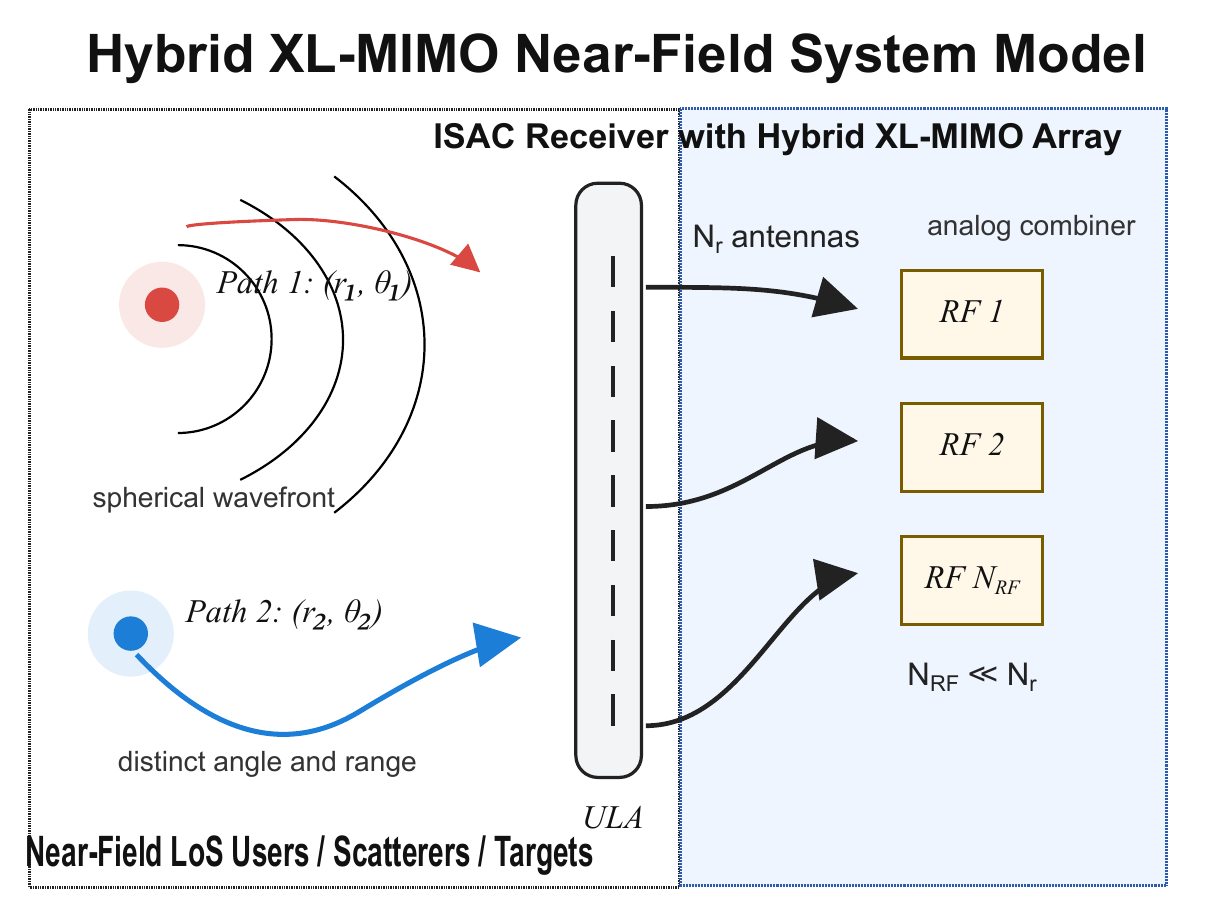}
    \caption{Hybrid XL-MIMO near-field system model. A base station with many antennas but few RF chains acquires compressed pilot measurements through analog combining, while each propagation path is parameterized by coupled angle and range.}
    \label{fig:xlmimo}
\end{figure}
We consider the uplink hybrid XL-MIMO architecture in Fig.~\ref{fig:xlmimo}, where the base station employs a ULA with $N_r$ antennas, spacing $d=\tfrac{\lambda}{2}$, and only $N_{\mathrm{RF}}\!\ll\! N_r$ RF chains. In pilot slot $p$, the analog network applies a constant-modulus combining matrix $\mx B_p\in\C^{N_{\mathrm{RF}}\times N_r}$ with $|B_p(i,j)|=\tfrac{1}{\sqrt{N_r}}$ \cite{alkhateeb2014channel,cui2022channel}. For a channel vector $\mx h\in\C^{N_r}$, pilot $x_p$, and noise $\mx w_p\sim\mathcal{CN}(\mx 0,\sigma^2\mx I)$, the compressed observation is \(\mx y_p=\mx B_p\mx h\,x_p+\mx B_p\mx w_p .
\) Stacking $P_T$ pilot slots and taking unit pilots for simplicity gives
\begin{equation}
\mx y=\mx B\mx h+\mx w\in\C^M,\qquad M\triangleq P_TN_{\mathrm{RF}},
\label{eq:stackedmodel}
\end{equation}
where $\mx B=[\mx B_1^{\top},\ldots,\mx B_{P_T}^{\top}]^{\top}\in\C^{M\times N_r}$ and $\mx w$ is the effective stacked noise. After pre-whitening, \eqref{eq:stackedmodel} remains the effective hybrid measurement model \cite{daei2026convexity}.
The model is narrowband and assumes that the signal bandwidth \(B_{\rm sig}\) is small relative to the inverse of the maximum propagation-delay variation across the array, i.e.,
\(
B_{\rm sig}\tau_{\max}\ll 1
\),
where
\(
\tau_{\max}\triangleq \max_{n,r,\theta}\tfrac{|r^{(n)}(r,\theta)-r|}{c_0}
\)
over the considered range-angle region, and \(c_0\) is the speed of light. Wideband spatial effects and beam split are outside the present scope.
We adopt a sparse near-field channel model with $L$ number of paths as follows: \(\mx h=\sum_{\ell=1}^{L} c_\ell\,\mx a_{\rm NF}(r_\ell,\thvar_\ell),
\)
where $c_\ell\in\C$ and $(r_\ell,\thvar_\ell)$ denote the gain, range, and azimuth of path $\ell$. The exact spherical-wave steering entry at antenna index $n\in\{0,\ldots,N_r-1\}$ is \(a_{\rm NF}(r,\thvar)[n]=\exp\!\left(-jk_{\lambda}\big(r^{(n)}-r\big)\right),
\) with \(r^{(n)}=\sqrt{r^2+(nd)^2-2r(nd)\cos\thvar},
\qquad k_{\lambda}\triangleq\tfrac{2\pi}{\lambda}.\)
The objective is to recover the continuous parameters $\{(r_\ell,\thvar_\ell,c_\ell)\}_{\ell=1}^{L}$ from the compressed hybrid measurements in \eqref{eq:stackedmodel}.
\section{Proposed Approach}
\label{sec:proposed}

The main difficulty in near-field estimation is that the spherical-wave steering vector is nonlinear in both range and angle.
To expose exploitable structure, we adopt the standard second-order Fresnel approximation and apply the Jacobi-Anger identity $e^{jz\cos\psi}=\sum_{m\in\mathbb Z} j^m J_m(z)e^{jm\psi},
$ which yields the truncated harmonic expansion
\begin{equation}
 a_{\rm NF}(r,\thvar)[n]
 \approx
 \sum_{\ell=-I_1}^{I_1}\sum_{q=-I_2}^{I_2}
 j^{\ell+q}J_{\ell}(k_{\lambda} nd)\,F_{n,q}(\tfrac{1}{r})\,e^{j(\ell+2q)\thvar},
 \label{eq:ja_main_prop}
\end{equation}
where $F_{n,q}(x)\triangleq e^{-j\alpha_n x}J_q(\alpha_n x), \alpha_n \triangleq \tfrac{k_{\lambda}(nd)^2}{4}$ and $I_1, I_2$ are the truncation levels.
Here, $\ell$ captures the harmonics associated with the linear angular term, while $q$ captures the additional harmonics induced by curvature. Thus, near-field curvature does not remove harmonic structure; it enriches it, preserving a hidden Vandermonde core in angle while shifting the range dependence into the coefficient functions $F_{n,q}(\tfrac{1}{r})$.

\subsection{Continuous transform-domain lifting}
Since $F_{n,q}$ depends on distance through $x=\tfrac{1}{r}$, inverse range is the natural coordinate in which to model curvature. We therefore define the periodic map \(\uvar(r)
\triangleq
2\pi\,\tfrac{\tfrac{1}{r}-x_{\min}}{x_{\max}-x_{\min}},
\qquad
x_{\min}\triangleq\tfrac{1}{r_{\max}},\;\;x_{\max}\triangleq\tfrac{1}{r_{\min}},\)
and approximate the coefficient functions by a short Fourier model in $\uvar$,
\begin{equation}
F_{n,q}(x)\approx \sum_{k=-K_u}^{K_u} a_{n,q}[k]\,e^{jk\uvar(x)}.
\label{eq:ufourier_prop}
\end{equation}
This keeps range continuous through $\uvar(r)$ while imposing only a finite transform-domain bandwidth.
Define the continuous steering vectors \(\bs \phi_{\uvar}(\uvar)
\triangleq
\big[e^{-jK_u\uvar},\ldots,e^{jK_u\uvar}\big]^{\mathsf T}\in\C^{N_u},
\label{eq:phi_u_prop}\\
\mx v(\thvar)
\triangleq
\big[e^{-jI_{\rm off}\thvar},\ldots,e^{jI_{\rm off}\thvar}\big]^{\mathsf T}\in\C^{N_b},\)
with $I_{\rm off}=I_1+2I_2$, $N_u=2K_u+1$, and $N_b=2I_{\rm off}+1$.
The following proposition provides the key structural bridge between the physical near-field model and the lifted linear inverse problem.
\begin{prop}[Continuous transformed-domain factorization]
\label{prop:factorization_prop}
Let $k_{\thvar}=\ell+2q$ denote the composite angular harmonic in \eqref{eq:ja_main_prop}, and let $a_{n,q}[k]$ be the coefficients in \eqref{eq:ufourier_prop}. Then, for every antenna index $n$, there exists a deterministic matrix $\bs\Phi_n\in\C^{N_u\times N_b}$ such that
\begin{equation}
a_{\rm NF}(r,\thvar)[n]
\approx
\ip{\bs \Phi_n}{\phi_{\uvar}(\uvar(r))\,v(\thvar)^{\mathsf T}}.
\label{eq:phi_factorization_prop}
\end{equation}
One admissible construction is \(\bs \Phi_n[k,m]
=
\sum_{\substack{|\ell|\le I_1,\;|q|\le I_2\\ \ell+2q=k_{\thvar}(m)}}
j^{\ell+q}J_\ell(k_{\lambda}nd)\,a_{n,q}[k],\)
where $k_{\thvar}(m)\in\{-I_{\rm off},\ldots,I_{\rm off}\}$ indexes the $m$th angular harmonic.
\end{prop}
Proposition~\ref{prop:factorization_prop} shows that each continuous range-angle pair $(r,\thvar)$ is embedded as a structured rank-one atom
\(
\phi_{\uvar}(\uvar(r))\,v(\thvar)^{\mathsf T}
\)
in a higher-dimensional lifted space, and that the steering vector is represented as the action of a linear operator on this atom. For $L$ paths, define the lifted matrix
\begin{equation}
\mx X
\triangleq
\sum_{\ell=1}^{L} c_\ell\,
\phi_{\uvar}(\uvar_\ell)\,v(\thvar_\ell)^{\mathsf T}\in\C^{N_u\times N_b},
\label{eq:Xdef_prop}
\end{equation}
so that $h[n]\approx \ip{\bs \Phi_n}{\mx X}.
$ In practice, the quality of \eqref{eq:ufourier_prop} depends on how well $F_{n,q}(x)$ is approximated over $x\in[x_{\min},x_{\max}]$. A global low-order Fourier fit is often insufficient over a wide range interval, so we adopt a panelized weighted least-squares fit on overlapping inverse-range panels. For panel $p$, the local coefficients are obtained from \(\min_{a}\;
\sum_i \omega_i^{(p)}
\left|
F_{n,q}(x_i^{(p)})
-
\sum_{k=-K_{\rm loc}}^{K_{\rm loc}} a[k]e^{jk\uvar_i^{(p)}}
\right|^2
+\mu\|a\|_2^2,\)
where $r_i=\tfrac{1}{x_i}$ and $\omega_i^{(p)}\propto r_i^2$. This compensates for the nonuniform physical emphasis induced by uniform sampling in $\uvar$, and the local panel fits are blended and projected onto the global order-$K_u$ basis to obtain the coefficients $a_{n,q}[k]$. This panelized fit is what makes the inverse-range representation both compact and stable over a wide distance interval.

\subsection{Hybrid measurements and 2D atomic norm recovery}
Substituting \eqref{eq:phi_factorization_prop} into the hybrid observation model \eqref{eq:stackedmodel} gives
\begin{equation}
y[m]=\ip{\bs\Psi_m}{\mx X}+\epsilon[m],
\qquad
\bs\Psi_m\triangleq\sum_{n=0}^{N_r-1}B[m,n]\bs\Phi_n,
\label{eq:hybrid_lifted_prop}
\end{equation}
where $\mx X$ is the lifted matrix in \eqref{eq:Xdef_prop}. Equivalently, with $z=\vecop(\mx X)$ and sensing matrix $\mx B'\in\C^{M\times N_uN_b}$ formed from the rows $\vecop(\mx\Psi_m)^{\mathsf H}$, $\mx y=\mx B' \mx z+\bs{\epsilon}.
$ Define the continuous atomic set \(\mathcal A
\triangleq
\left\{
\bs \phi_{\uvar}(\uvar)\,\mx v(\thvar)^{\mathsf T}:\;
\uvar\in[0,2\pi),\;\thvar\in[0,\pi)
\right\},\) with associated atomic norm \(\|\mx X\|_{\mathcal A}
=
\inf\left\{
\sum_\ell |c_\ell|:\;
\mx X=\sum_\ell c_\ell\bs \phi_{\uvar}(\uvar_\ell)\mx v(\thvar_\ell)^{\mathsf H}
\right\}.\)
Recovery is posed as
\begin{equation}
\min_{\mx Z\in\C^{N_u\times N_b}}
\|\mx Z\|_{\mathcal A}
\quad
\text{s.t.}
\quad
\|\mx y-\mx B'\mx z\|_2\le \eta.
\label{eq:primal_prop}
\end{equation}
where $\eta$ is an upper-bound for $\|\bs \epsilon\|_2$. 
To express the atomic norm as a tractable semidefinite program, we propose the following problem which can be solved using standard convex solvers.
For 2D complex exponentials on a rectangular support, this norm admits the standard block-Toeplitz semidefinite lift \cite{tang2013compressed,candes2014towards,chandrasekaran2012convex,daei2025timely,bayat2020separating,safari2021off,dumitrescu2017positive,valiulahi2019two,razavikia2023off}. Writing $\mx z=\vecop(\mx Z)$,
\begin{align}
\min_{V,z,t}\;&
\tfrac{1}{2N_uN_b}\tr\big(\mathcal T_{\rm 2D}(V)\big)+\tfrac{t}{2}
\label{eq:sdp_prop}\\
\text{s.t. }&
\begin{bmatrix}
\mathcal T_{\rm 2D}(\mx V) & \mx z\\
\mx z^{\mathsf H} & t
\end{bmatrix}\succeq 0,
\qquad
\|\mx y-\mx B' \mx z\|_2\le \eta.
\nonumber
\end{align}
Here, the 2D block-Toeplitz lifting operator
\(
\mathcal T_{\rm 2D}:\C^{(2N_u-1)\times(2N_b-1)}\rightarrow \C^{N_uN_b\times N_uN_b},
\) maps the lag array $V$ to the Hermitian moment matrix indexed by the $u$- and $\theta$-lags. More precisely, with the $u$-fast vectorization rule
\(
r=(i_b-1)N_u+i_u,\qquad c=(j_b-1)N_u+j_u,
\)
for $i_u,j_u\in\{1,\dots,N_u\}$ and $i_b,j_b\in\{1,\dots,N_b\}$, we define $\big[\mathcal T_{\rm 2D}(V)\big]_{r,c}
\triangleq
V(i_u-j_u+N_u,\; i_b-j_b+N_b).$ Thus, $\mathcal T_{\rm 2D}(\mx V)$ is Toeplitz across the $u$ dimension and block-Toeplitz across the $\theta$ dimension. Regarding the Computational Complexity, let \(N=N_uN_b\). The SDP in \eqref{eq:sdp_prop} has a main PSD block of size \(N+1\); hence a generic interior-point solver has per-iteration complexity approximately
\(
\mathcal O(n_v^2N^2+n_vN^3+MN),
\)
with \(n_v=\mathcal O(N)\) scalar variables up to constants. Thus, the SDP is suitable as a convex gridless benchmark for moderate lifted dimensions, while larger arrays would require first-order or structure-exploiting solvers.
Let $\mx q\in \mathbb C^{M}$ be the dual vector corresponding to \eqref{eq:primal_prop}. Since the lifted model is $\mx y=\mx B'\mx z+\bs{\epsilon}$, the associated dual polynomial is \(Q(\uvar,\thvar)
\triangleq
\ip{\mx B'^{\mathsf H}\mx q}
{\vecop\!\left(\bs \phi_{\uvar}(\uvar)\mx v(\thvar)^{\mathsf T}\right)} .\)
Its peaks provide a certificate-based localization rule in the continuous transformed domain. Under standard dual-certificate conditions, \( |Q(\uvar,\thvar)|=1 \) at active support locations and \( |Q(\uvar,\thvar)|<1 \) away from them \cite{fernandez2016super,dumitrescu2017positive,daei2025timely}. The range estimate $\hat r$ is then obtained from the inverse map of $\hat \uvar$.
Given the recovered support $\widehat{\mathcal S}=\{(\hat u_\ell,\hat\theta_\ell)\}_{\ell=1}^{\hat L}$, the corresponding complex amplitudes are obtained by least-squares refinement. Defining $\hat {\mx A}_\ell\triangleq \bs \phi_{\uvar}(\hat u_\ell)\mx v(\hat\theta_\ell)^{\mathsf T}$ and $G[m,\ell]\triangleq \ip{\bs\Psi_m}{\hat {\mx A}_\ell},$ the gain vector is estimated as \(\hat{\mathbf c}
=
\arg\min_{{\mx c}\in\mathbb{C}^{\hat L\times 1}}\|\mathbf y-\mx G\mathbf c\|_2^2
=
(\mx G^{\mathsf H}\mx G)^{-1}\mx G^{\mathsf H}\mathbf y,\)
provided that $\mx G$ has full column rank.

\section{Numerical Results}
\begin{figure}[t]
\centering
\begin{minipage}[t]{0.3\columnwidth}
\vspace{0pt}
\centering
\scriptsize
\setlength{\tabcolsep}{.8pt}
\begin{tabular}{lll}
\toprule
Param. & P1 & P2 \\
\midrule
$\theta$        & $0.8749$ & $1.9866$ \\
$\hat{\theta}$  & $0.8770$ & $1.9897$ \\
$r$ [m]         & $3.4172$ & $0.8560$ \\
$\hat r$ [m]    & $3.4128$ & $0.8550$ \\
$u$             & $0.2499$ & $1.9873$ \\
$\hat u$        & $0.2507$ & $1.9899$ \\
$c_{\ell}$      & $1.2968+0.6096i$ & $0.3802-1.5972i$ \\
$\hat c_{\ell}$ & $1.2912+0.6169i$ & $0.3898-1.5956i$ \\
\bottomrule
\end{tabular}
\end{minipage}\hfill
\begin{minipage}[t]{0.45\columnwidth}
\vspace{0pt}
\centering
\includegraphics[width=\linewidth]{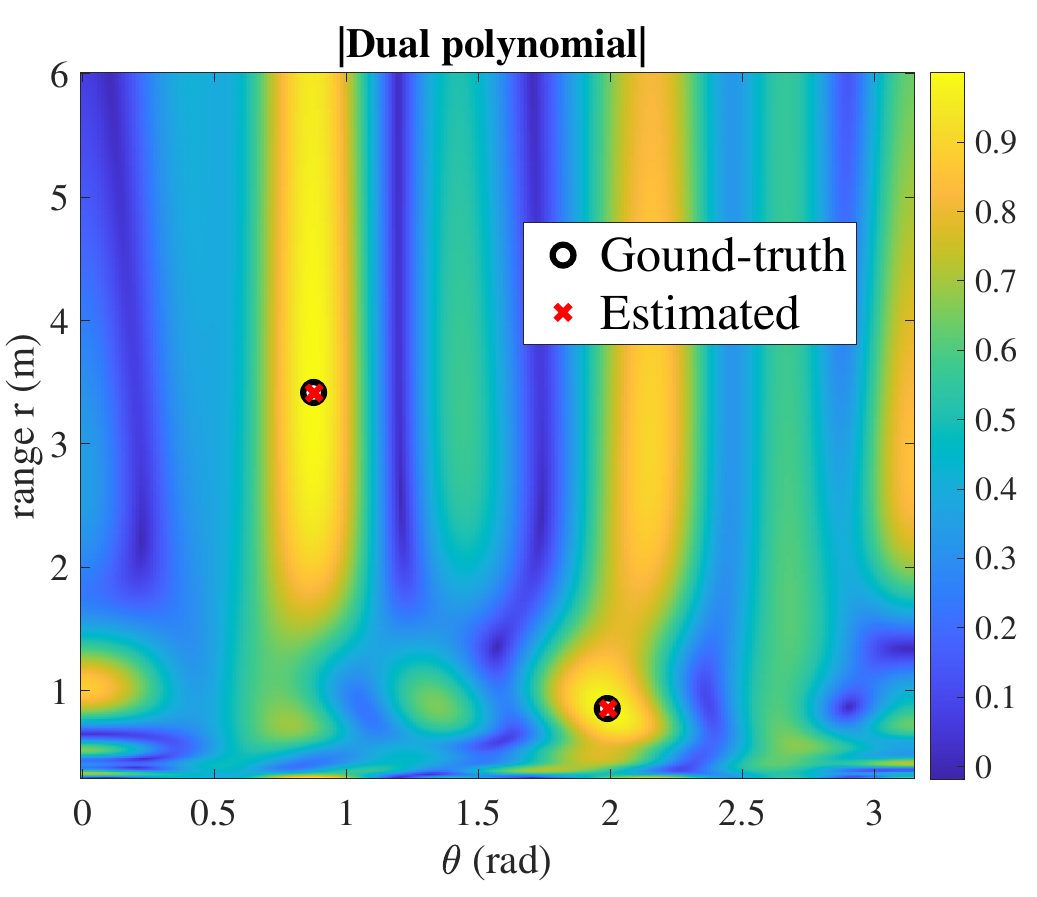}
\end{minipage}
\caption{Dual-polynomial localization and corresponding parameter estimates for the representative two-path experiment. The right panel shows the magnitude of the dual polynomial originally computed on a uniform $(u,\theta)$ grid and then interpolated onto a uniform $(r,\theta)$ grid for visualization. The markers indicate the true and recovered scatterer locations, while the left panel reports the corresponding true and recovered parameters.}
\label{fig:dual_pol}
\end{figure}

We performed numerical experiments to verify our method. The BS employs a hybrid XL-MIMO array with $N_r=64$ antennas at carrier frequency $f_c=100$ GHz and inter-element spacing $d=\tfrac{\lambda}{2}$. The hybrid front-end provides $M\triangleq P_TN_{\mathrm{RF}}=20$ effective compressed pilot measurements, modeled by a constant-modulus combining matrix $\mx B\in\C^{20\times 64}$. The continuous transformed-domain model uses truncation orders $I_1=20$ and $I_2=1$ for the angular and curvature harmonics, respectively, together with global inverse-range order $K_u=2$ and local panel order $K_{\rm loc}=2$. This yields a lifted matrix dimension of $N_u\times N_b = 5\times 45$.
The range interval is $[r_{\min},r_{\max}] = [0.1,6]$ m, and the inverse-range approximation is constructed over $P=4$ overlapping panels in $x=\tfrac{1}{r}$. The scene contains $L=2$ paths with unit gains, where the ranges are drawn uniformly over $[0.1,6]$ m and the paths are separated in the transformed $(\uvar,\thvar)$ domain. The reported experiment is noiseless, namely $\mx y=\mx B\mx h+ \bs \epsilon$ where $\bs \epsilon$ denotes the approximation error. The same framework extends directly to noisy measurements by incorporating an additional noise term in the observation model. The ground-truth range-angle parameters with their corresponding estimates obtained from our approach are shown in Figure \ref{fig:dual_pol}.
% For the representative run used here, the true path parameters were
% \begin{align*}
% (\thvar_1,r_1,\uvar_1) &= (1.9638\,\text{rad},\;5.3936\,\text{m},\;0.0120),\\
% (\thvar_2,r_2,\uvar_2) &= (1.1609\,\text{rad},\;0.1220\,\text{m},\;5.1296).
% \end{align*}
The SDPT3/CVX implementation solved \eqref{eq:sdp_prop} and recovered both paths to the resolution of the dual-polynomial peak search. The least-squares amplitude refinement then accurately estimated the complex-valued gains \(\hat c_{\ell}\), as shown in Fig.~\ref{fig:dual_pol}. This is a demanding scene because both scatterers lie in the near-field region of the BS, yet both are recovered from only \(M=20\) hybrid-compressed observations.
% \begin{table}[t]
% \caption{Parameters Used in Numerical Results}
% \label{tab:sim}
% \centering
% \small
% \begin{tabular}{>{\raggedright\arraybackslash}p{1.55cm} p{5.15cm}}
% \toprule
% Parameter & Value \\
% \midrule
% Hybrid BS & $N_r=64$ antennas, $M=P_TN_{\mathrm{RF}}=20$ effective compressed measurements \\
% Array/carrier & $f_c=100$ GHz, $d=\lambda/2$ \\
% Range & $[r_{\min},r_{\max}] = [0.1,6]$ m \\
% Orders & $I_1=15$, $I_2=1$, $K_u=2$, $K_{\rm loc}=2$ \\
% Lifted dim. & $N_u\times N_b = 5\times 35$ \\
% Panels & $P=4$ overlapping panels in $x=\tfrac{1}{r}$ \\
% Scene & $L=2$ unit-gain paths with uniform-$r$ draw and transformed-domain separation \\
% Recovery & noiseless surrogate-consistent, $\eta=0$ \\
% \bottomrule
% \end{tabular}
% \end{table}
The main message is not only that the support is accurately recovered in this representative noiseless experiment, but that the hybrid XL-MIMO front-end can be handled without any range and angle discretization in the recovery stage. The algorithm first converts the near-field manifold into a compact continuous transform-domain model, then applies a convex 2D super-resolution program directly to the hybrid-compressed measurements.

\section{Conclusion}
This paper introduced a new way to think about near-field super-resolution under hybrid XL-MIMO measurements: not as a gridded range-angle search problem, but as a continuous inverse problem in a geometry-aligned transformed domain. By reparameterizing distance through inverse range, exposing a Bessel-Vandermonde structure in the near-field manifold, and lifting each continuous range-angle pair to a structured rank-one atom, we converted a nonlinear spherical-wave model into a linear recovery problem amenable to convex optimization. The resulting inverse-range atomic norm framework remains continuous in both variables, preserves linearity under hybrid combining, and admits certificate-based localization through a block multidimensional Toeplitz lift.
Beyond the specific formulation, the broader message is that near-field curvature does not have to be handled through ever finer discretization. When represented in the right coordinates, it reveals structure that can be exploited directly. Numerical results demonstrate that the proposed framework is not only conceptually appealing but also practically effective. These results position inverse-range atomic norms as a promising foundation for gridless near-field sensing and channel estimation in hybrid XL-MIMO and ISAC systems. Extending the approximation analysis, recovery guarantees, and wideband formulation to exact spherical-wave measurements over wider Fresnel regimes is a natural future step.
\appendix

Substituting the inverse-range Fourier model \eqref{eq:ufourier_prop} into the truncated harmonic expansion \eqref{eq:ja_main_prop},
% \begin{equation}
% F_{n,q}(\tfrac{1}{r})\approx \sum_{k=-K_u}^{K_u} a_{n,q}[k]\,e^{jk\uvar(r)},
% \label{eq:pf2}
% \end{equation}
gives
\begin{align}
\scalebox{.8}{$a_{\rm NF}(r,\thvar)[n]
\approx
\sum_{k=-K_u}^{K_u}\sum_{|\ell|\le I_1}\sum_{|q|\le I_2}
j^{\ell+q}J_\ell(k_{\lambda}nd)\,a_{n,q}[k]\,
e^{jk\uvar(r)}e^{j(\ell+2q)\thvar}.$}
\label{eq:pf3}
\end{align}

Now group together all terms having the same composite angular harmonic
\(
k_{\thvar}=\ell+2q.
\)
Since $|\ell|\le I_1$ and $|q|\le I_2$, we have $k_{\thvar}\in\{-I_{\rm off},\ldots,I_{\rm off}\}$ with $I_{\rm off}=I_1+2I_2$. Let $k_{\thvar}(m)$ denote the $m$th harmonic index, $m=1,\ldots,N_b$, and define \(\bs \Phi_n[k,m]
\triangleq
\sum_{\substack{|\ell|\le I_1,\;|q|\le I_2\\ \ell+2q=k_{\thvar}(m)}}
j^{\ell+q}J_\ell(k_{\lambda}nd)\,a_{n,q}[k].\)
Then \eqref{eq:pf3} becomes
\begin{equation}
a_{\rm NF}(r,\thvar)[n]
\approx
\sum_{k=-K_u}^{K_u}\sum_{m=1}^{N_b}
\bs \Phi_n[k,m]\,
e^{jk\uvar(r)}e^{jk_{\thvar}(m)\thvar}.
\label{eq:pf5}
\end{equation}

By the definitions of the transformed-domain steering vectors,
\(\phi_{\uvar}(\uvar(r))[k]=e^{jk\uvar(r)}\) and
\(v(\thvar)[m]=e^{jk_{\thvar}(m)\thvar}\). The $(k,m)$th entry of the rank-one atom
\(\phi_{\uvar}(\uvar(r))\,v(\thvar)^{\mathsf T}\)
is exactly \(e^{jk\uvar(r)}e^{jk_{\thvar}(m)\thvar}\) under the adopted indexing convention.
Hence \eqref{eq:pf5} can be written as
\begin{align}
&a_{\rm NF}(r,\thvar)[n]
\approx
\sum_{k=-K_u}^{K_u}\sum_{m=1}^{N_b}
\bs \Phi_n[k,m]\,
\big[\bs \phi_{\uvar}(\uvar(r))\mx v(\thvar)^{\mathsf T}\big]_{k,m}\nonumber\\
&=
\ip{\bs \Phi_n}{\bs \phi_{\uvar}(\uvar(r))v\mx (\thvar)^{\mathsf T}},
\end{align}
which proves the claim of the proposition.

\bibliographystyle{IEEEtran}
\bibliography{spawc_refs}
\end{document}